\documentclass{PoS}

\usepackage{graphicx}
\usepackage{amsmath}
\usepackage{amssymb}
\usepackage{listings} 

\newcommand{\asb}{\bar{\alpha}_s}

\newcommand{\stringa}{\ttfamily\lstinline}
\def\cod#1{{\stringa!#1!}}
\newcommand{\vsb}{\vspace{-0.175cm}}

\title{Inclusive four-jet production: a study of 
       Multi-Regge kinematics and BFKL observables}

\ShortTitle{Inclusive four-jet production: a study of 
            Multi-Regge kinematics and BFKL observables}
            
\author{Francesco Caporale \\
        Instituto de Física Teórica UAM/CSIC, 
        Nicolás Cabrera 15, Madrid, Spain \\
        Universidad Autónoma de Madrid, 28049 Madrid, Spain
        E-mail: 
        \email{francesco.caporale@uam.es
}}

\author{\speaker{Francesco Giovanni Celiberto} \\
        Dipartimento di Fisica, Università della Calabria
        and Istituto Nazionale di Fisica Nucleare, 
        Gruppo Collegato di Cosenza,
        Arcavacata di Rende, 87036 Cosenza, Italy, \\
        Instituto de Física Teórica UAM/CSIC, 
        Nicolás Cabrera 15, Madrid, Spain \\
        Universidad Autónoma de Madrid, 28049 Madrid, Spain
        E-mail: 
        \email{francescogiovanni.celiberto@fis.unical.it}}

\author{Grigorios Chachamis \\
        Instituto de Física Teórica UAM/CSIC, 
        Nicolás Cabrera 15, Madrid, Spain \\
        Universidad Autónoma de Madrid, 28049 Madrid, Spain \\
        \email{chachamis@gmail.com
}}
        
\author{Agustín Sabio Vera \\
        Instituto de Física Teórica UAM/CSIC, 
        Nicolás Cabrera 15, Madrid, Spain \\
        Universidad Autónoma de Madrid, 28049 Madrid, Spain
        E-mail: 
        \email{a.sabio.vera@gmail.com}}

\abstract{A study of differential cross sections 
          for the production of four jets in 
          multi-Regge kinematics is presented, 
          the main focus lying on azimuthal angle
          dependences. The theoretical setup consists 
          in the jet production from a single BFKL
          ladder with a convolution of three 
          BFKL Green functions, where two forward/backward
          jets are always tagged in the final state. 
          Furthermore, the tagging of two further
          jets in more central regions of the detectors 
          with a relative separation in rapidity
          from each other is requested. It is found, 
          as result, that the dependence on the
          transverse momenta and the rapidities 
          of the two central jets can be considered as a
          distinct signal of the onset of BFKL dynamics.
          }

\FullConference{XXIV International Workshop on 
                Deep-Inelastic Scattering 
                and Related Subjects\\
		11-15 April, 2016\\
		DESY Hamburg, Germany}

\begin{document}

\section{Introduction}

The study of semi-hard processes 
in the high-energy (Regge) limit
is an ultimate testfield for pQCD, 
the Large Hadron Collider (LHC) 
providing with an abundance of data. 
Multi-Regge kinematics (MRK),
which prescribes a strong ordering in rapidity 
for the final state objects,
is the key ingredient for the study 
of multi-jet production at LHC energies.
In this kinematical regime, 
the common basis for the perturbative description 
of semi-hard processes 
is the Balitsky-Fadin-Kuraev-Lipatov 
(BFKL) approach, at leading (LL) 
\cite{Lipatov:1985uk,Balitsky:1978ic,
Kuraev:1977fs,Kuraev:1976ge,Lipatov:1976zz,Fadin:1975cb} 
and next-to-leading 
(NLL)~\cite{Fadin:1998py,Ciafaloni:1998gs} accuracy,
which offers a resummation of those enhanced terms 
in MRK in regions of phase space where a fixed order 
calculation might not be enough. 
This formalism has been successfully applied 
to lepton-hadron Deep Inelastic Scattering at HERA 
(see, {\it e.g.}~\cite{Hentschinski:2012kr,
Hentschinski:2013id}) in order to study quite inclusive 
processes which are not that suitable though 
to discriminating between BFKL dynamics 
and other resummations. 
At the LHC, however, it is possible 
to investigate processes with much more exclusive 
final states which could, in principle, 
be only described by the BFKL framework, allowing us 
to disentangle the applicability region of the approach. 
In these last years, the most investigated process was the 
Mueller-Navelet jet production~\cite{Mueller:1986ey}.
Interesting observables associated to this reaction are 
the azimuthal correlation momenta which, however, 
are strongly affected by collinear contaminations.
Therefore, new observables independent 
from the conformal contribution
were proposed in~\cite{Vera:2006un,Vera:2007kn} 
and calculated at NLL 
in~\cite{Ducloue:2013bva,Caporale:2014gpa,
Celiberto:2015yba,Celiberto:2016ygs,
Ciesielski:2014dfa,Angioni:2011wj,Chachamis:2015crx},  
showing a very good agreement 
with experimental data at the LHC. 
Nevertheless, Mueller-Navelet configurations 
are still too inclusive to perform MRK precision studies. 
For this reason a new study that demands the tagging 
of a third, central in rapidity, jet 
within the usual Mueller-Navelet configuration
was proposed~\cite{Caporale:2015vya,Caporale:2016soq}.
Another interesting and novel possibility, 
the detection of two charged light hadrons:
$\pi^{\pm}$, $K^{\pm}$, $p$, $\bar p$ 
having high transverse momenta and 
separated by a large interval of rapidity,
together with an undetected soft-gluon radiaton emission, 
was suggested in~\cite{Ivanov:2012iv} 
and studied in~\cite{Celiberto:2016hae}.
In the present work we follow the course taken 
in~\cite{Caporale:2015vya,Caporale:2016soq}, 
by allowing the production of a second central jet, 
which makes it possible to define new, generalized 
differential distributions~\cite{Caporale:2015int}
in the transverse momenta, azimuthal angles 
and rapidities of the two central jets, 
for fixed values of the four momenta of the forward jets.

\section{Inclusive four-jet production}

We study events with 
two forward/backward jets together with two more central 
jets, all of them well separated in rapidity 
from each other, making use of the BFKL formalism.  
The two tagged forward/backward jets $A$ and $B$ 
have transverse momentum $\vec{k}_{A,B}$, azimuthal angle 
$\vartheta_{A,B}$ and rapidity $Y_{A,B}$, 
while the pair of tagged more central jets 
are characterized,  respectively, 
by $\vec{k}_{1,2}$, $\vartheta_{1,2}$ and $y_{1,2}$. 
The differential cross section is
\begin{align}\label{d6sigma}
 &
 \frac{d^6\sigma^{\rm 4-jet} 
 \left(\vec{k_A},\vec{k_B},Y_A-Y_B\right)} 
      {d^2\vec{k_1} dy_1 d^2\vec{k_2} dy_2}
 =
 \frac{\asb^2}{\pi^2 k_1^2 k_2^2}
 \int d^2\vec{p_A} \int d^2\vec{p_B} \int d^2\vec{p_1} 
 \int d^2\vec{p_2}
 \delta^{(2)}\left(\vec{p_A}+\vec{k_1}-\vec{p_1}\right)
 \\ & \nonumber \hspace{1.25cm}
 \delta^{(2)}\left(\vec{p_B}-\vec{k_2}-\vec{p_2}\right)
 \varphi\left(\vec{k_A},\vec{p_A},Y_A-y_1\right)
 \varphi\left(\vec{p_1},\vec{p_2},y_1-y_2\right)
 \varphi\left(\vec{p_B},\vec{k_B},y_2-Y_B\right).
\end{align}
Here we have introduced the rapidity ordering 
characteristic of MRK: $Y_A > y_1 > y_2 > Y_B$; 
and $k_1^2$, $k_2^2$ lie above 
the experimental resolution scale. 
$\varphi$ are BFKL gluon Green functions normalized 
to $ \varphi \left(\vec{p},\vec{q},0\right) 
= \delta^{(2)} \left(\vec{p} - \vec{q}\right)/(2 \pi)$ 
and $\bar{\alpha}_s = \alpha_s N_c/\pi$.
We start with the study 
of an observable similar to the usual Mueller-Navelet case 
such that we integrate over the azimuthal angles 
of the two central jets and over the difference 
in azimuthal angle between the two forward jets, 
$\Delta\theta = \vartheta_A - \vartheta_B - \pi$, 
considering so the mean value of the cosine 
\begin{equation}
\label{C0}
 \left\langle\cos(M(\vartheta_A - \vartheta_B - \pi))\right\rangle
 =
 \frac{\int_0^{2\pi} d\Delta\theta \cos(M\Delta\theta)
       \int_0^{2\pi} d\vartheta_1 \int_0^{2\pi} d\vartheta_2
       \frac{d^6\sigma^{\rm 4-jet}}
            {dk_1 dy_1 d\vartheta_1 dk_2 d\vartheta_2 dy_2}}
      {\int_0^{2\pi} d\Delta\theta 
       \int_0^{2\pi} d\vartheta_1 \int_0^{2\pi} d\vartheta_2
       \frac{d^6\sigma^{\rm 4-jet}}
            {dk_1 dy_1 d\vartheta_1 dk_2 d\vartheta_2 dy_2}}.
\end{equation}

Our next step now is to propose new observables, 
different from those characteristic of the Mueller-Navelet 
case though still related 
to azimuthal angle projections. We thus define
\begin{align}\label{projections_1} 
 \mathcal{C}_{MNL} = & 
 \int_0^{2\pi} d\vartheta_A \int_0^{2\pi} d\vartheta_B
 \int_0^{2\pi} d\vartheta_1 \int_0^{2\pi} d\vartheta_2
 \ 
 \cos\left(M\left(\vartheta_A-\vartheta_1-\pi\right)\right)
 \\ & \nonumber 
 \cos\left(N\left(\vartheta_1-\vartheta_2-\pi\right)\right)
 \cos\left(L\left(\vartheta_2-\vartheta_B-\pi\right)\right)
 \frac{d^6\sigma^{\rm 4-jet}\left(\vec{k_A},
                                  \vec{k_B},Y_A-Y_B\right)}
 {dk_1 dy_1 d\vartheta_1 dk_2 d\vartheta_2 dy_2}\,,
 \end{align}
In order to improve the perturbative stability 
of our predictions (see~\cite{Caporale:2013uva} 
for a related discussion) it is convenient to remove 
the contribution from the zero conformal spin 
(which corresponds to the index $n=0$ 
in Eq.~(\ref{projections_1})) by defining the ratios
\begin{equation}\label{R^mnl_pqr}
 \mathcal{R}^{MNL}_{PQR}
 =
 \frac{\left\langle\cos(M(\vartheta_A - \vartheta_1 - \pi))
                   \cos(N(\vartheta_1 - \vartheta_2 - \pi))
                   \cos(L(\vartheta_2 - \vartheta_B - \pi))
                   \right\rangle}
      {\left\langle\cos(P(\vartheta_A - \vartheta_1 - \pi))
                   \cos(Q(\vartheta_1 - \vartheta_2 - \pi))
                   \cos(R(\vartheta_2 - \vartheta_B - \pi))
                   \right\rangle}
\end{equation}
with integer $M,N,L,P,Q,R > 0$. 

We present now a numerical analysis done 
in two characteristic configurations 
for the transverse momenta of the forward jets, 
namely $ k_A \sim k_B $ and $ k_A < k_B $ 
(or equivalently $k_A > k_B$),
by fixing them to $\left( k_A, k_B \right)$ = $(40, 50)$  
and to $\left( k_A, k_B \right)$ = $(30, 60)$ GeV. 
We also fix the rapidities of the four tagged jets 
to the values $Y_A = 9$, $y_1 = 6$, $Y_2 = 3$, 
and $Y_B = 0$ whereas the two inner jets 
can have transverse momenta 
in the range $20 < k_{1,2} < 80$ GeV. 

\begin{figure}[t]
\centering
   \includegraphics[scale=0.60]{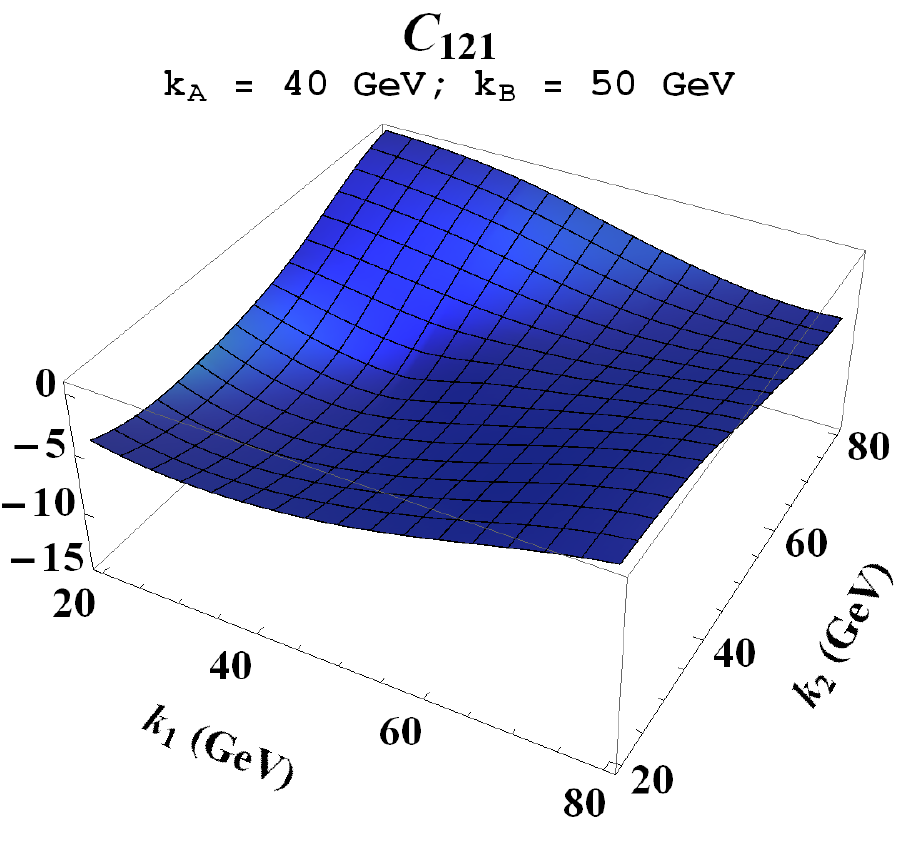}
   \includegraphics[scale=0.60]{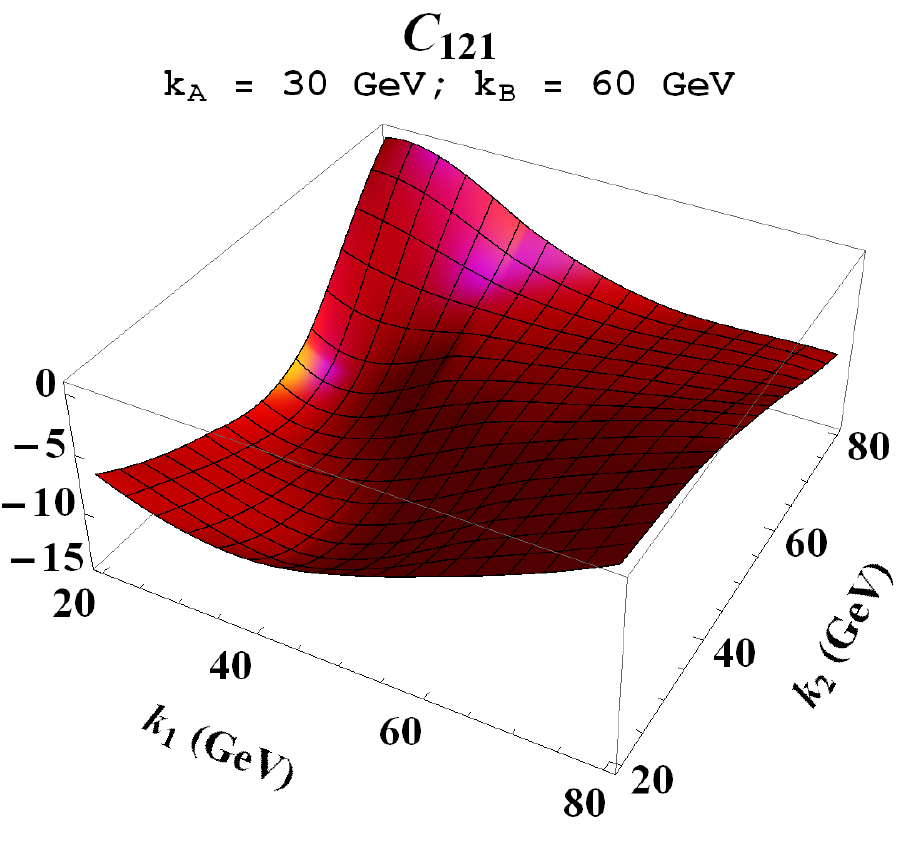}
   
   \includegraphics[scale=0.60]{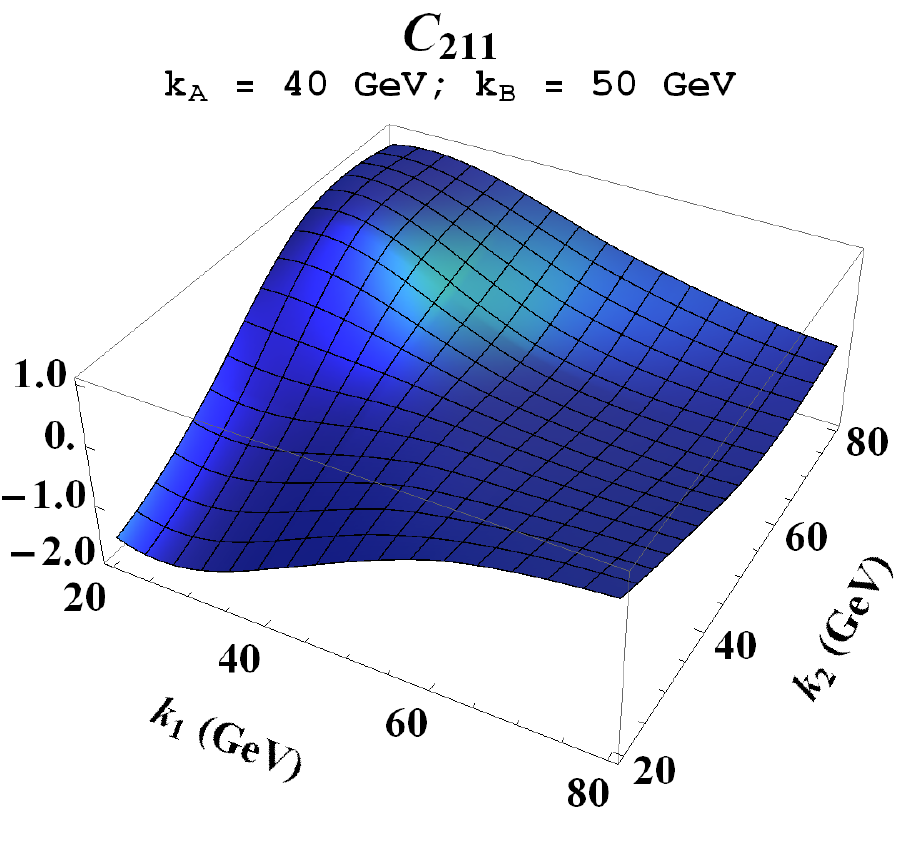}
   \includegraphics[scale=0.60]{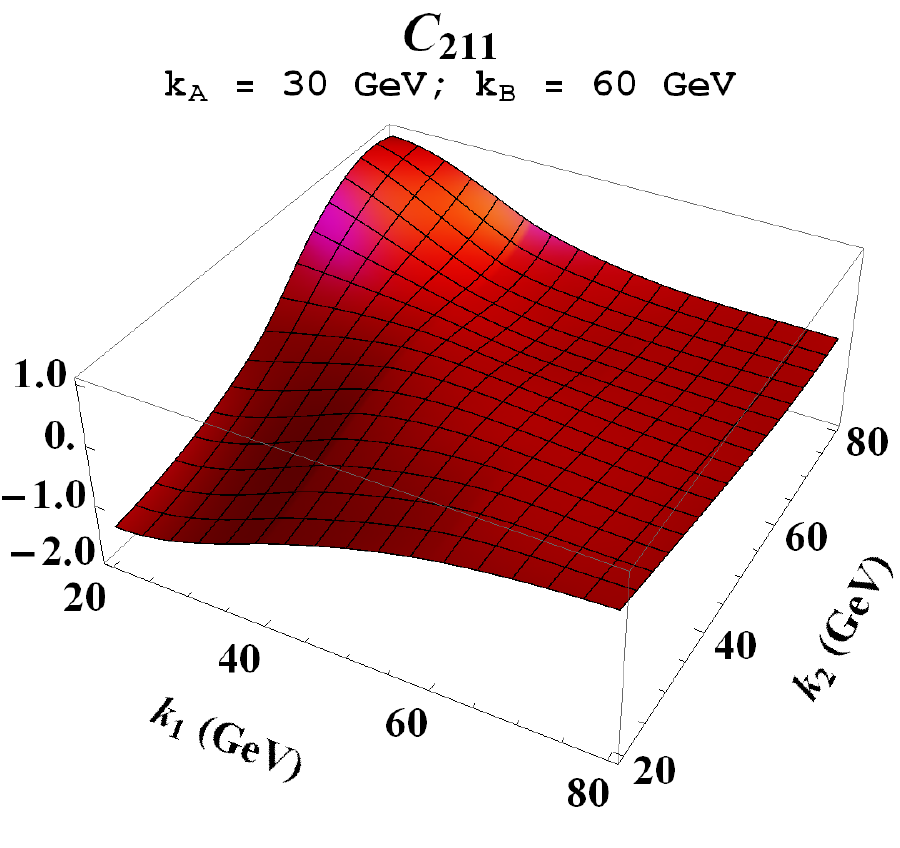}
\caption{\small $k_{1,2}$-dependence of the normalized 
${\cal C}_{121}$ and ${\cal C}_{211}$ 
for the two selected cases 
of forward jet transverse momenta $k_A$ and $k_B$.}
\label{C1nl}
\end{figure}

\begin{figure}[t]
\centering
   \includegraphics[scale=0.60]{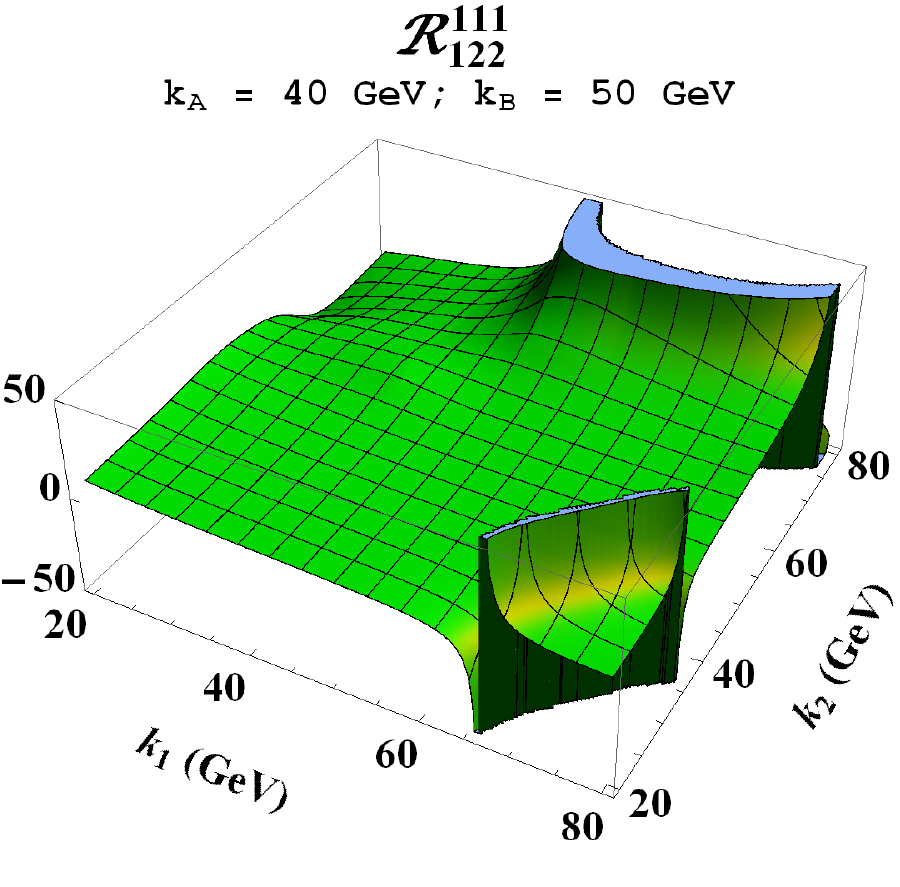}
   \includegraphics[scale=0.60]{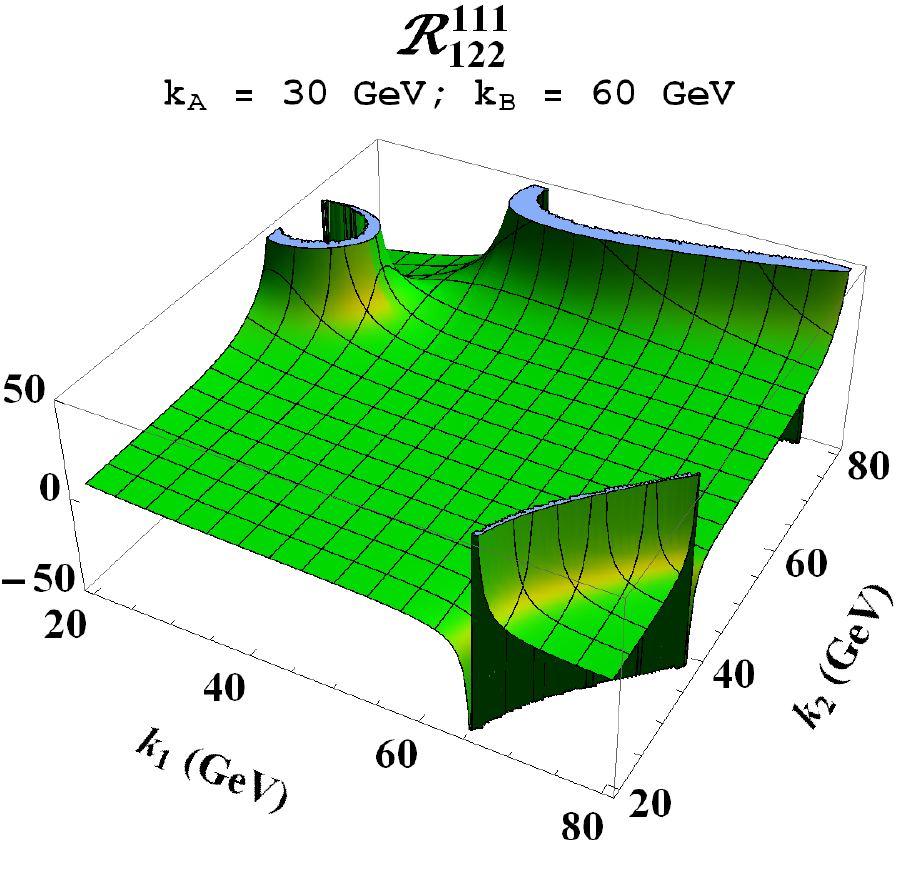}

   \includegraphics[scale=0.60]{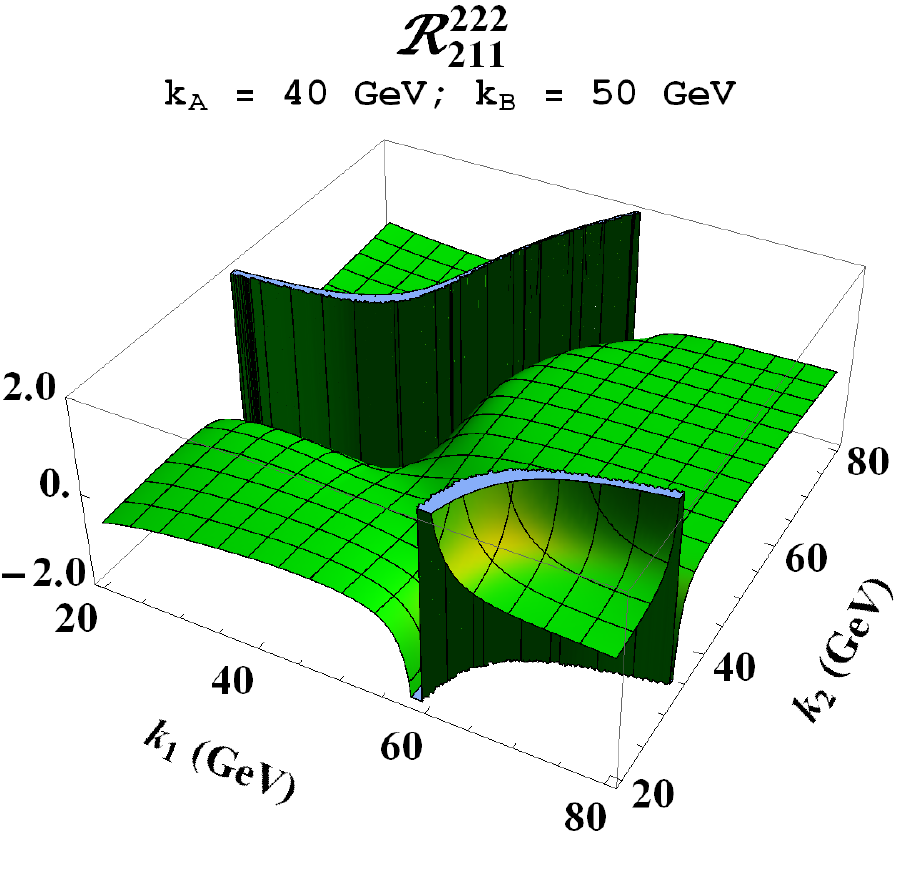}
   \includegraphics[scale=0.60]{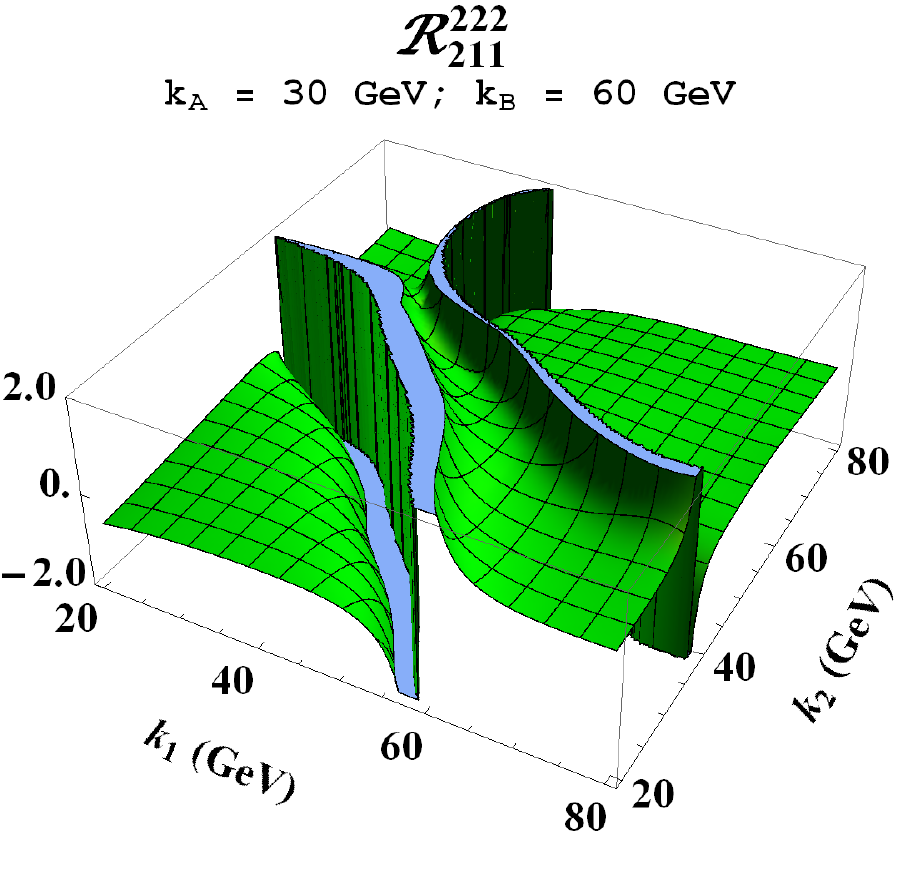}
\caption{\small $k_{1,2}$-dependence of 
$\mathcal{R}^{111}_{122}$ and $\mathcal{R}^{222}_{211}$ 
for the two selected cases of forward/backward jets  
transverse momenta $k_A$ and $k_B$.}
\label{fig:ratios_2}
\end{figure}

In Fig.~\ref{C1nl} we show the behaviour 
for the normalized
coefficients ${\cal C}_{121}$ and ${\cal C}_{211}$ 
after they are divided by their respective maximum.
We find that the distributions are quite similar 
for the two configurations here chosen 
($\left( k_A, k_B \right)$ = $(40, 50)$, $(30, 60)$ GeV).
Since these coefficients change sign 
in the $k_{1,2}$ range studied, 
it is clear that for the associated ratios 
$\mathcal{R}^{MNL}_{PQR}$ there will be some lines 
of singularities. We present the behaviour of 
$\mathcal{R}^{111}_{122}$ and 
$\mathcal{R}^{222}_{211}$ in Fig.~\ref{fig:ratios_2}. 
In this case the configurations 
$\left( k_A, k_B \right)$ = $(40, 50)$, $(30, 60)$ GeV 
behave quite differently. This is due to the variation 
of the position of the zeroes 
of those coefficients ${\cal C}_{MNL}$ 
chosen as denominators in these quantities. 
It would be very interesting to test if these singularity 
lines are present in any form in the LHC experimental data. 
In general, we notice a very weak dependence 
on variations of the rapidities $y_{1,2}$ 
of the more central jets 
for all the observables here presented. 

We have used both \textsc{Fortran} 
and \textsc{Mathematica} for the numerical computation
of the ratios $\mathcal{R}^{MNL}_{PQR}$. 
We made extensive use of the integration routine 
{\bf\cod{Vegas}}~\cite{VegasLepage:1978} 
as implemented in the {\bf\cod{Cuba}} 
library~\cite{Cuba:2005,ConcCuba:2015}.
Furthermore, we used the {\bf\cod{Quadpack}} 
library~\cite{Quadpack:book:1983}
and a slightly modified version of the 
{\bf\cod{Psi}}~\cite{RpsiCody:1973} routine.

It will be very important to 
compare against the BFKL Monte Carlo code 
{\bf\cod{BFKLex}}~\cite{Chachamis:2013rca,Caporale:2013bva,
Chachamis:2012qw,Chachamis:2012fk,Chachamis:2011nz,
Chachamis:2011rw,Chachamis:2015zzp}
as well as to calculate the same quantities
with other, more conventional, 
approaches~\cite{Bury:2015dla,
Nefedov:2013ywa,vanHameren:2012uj}
in order to gauge if they differ from our results. 
This includes those analysis 
where the four-jet predictions 
stem from two independent gluon 
ladders~\cite{Maciula:2015,Maciula:2014pla,
              Kutak:2016mik,Kutak:2016ukc}.

\section{Conclusions \& Outlook}

We have considered ratios of correlation functions 
of products of azimuthal angle difference cosines
in order to study four-jet production at hadron colliders. 
A single gluon ladder approach, 
with inclusive production of two forward/backward 
and two further, more central, tagged jets has been used. 
The dependence on the transverse momenta and rapidities 
of the two central jets represents a clear signal 
of BFKL dynamics. 
We continued and extended this study 
in~\cite{Caporale:2016xku} by giving 
more realistic predicions on the hadronc level
through the introduction of parton distribution functions. 
For future works, more accurate analyses are needed: 
higher order effects and study of different configurations 
for the rapidity range of the two central jets. 
It is also interesting to calculate our proposed observables 
using other approaches not based on the BFKL approach 
and to test how they can differ from our predictions. 
In view of all these considerations, we encourage 
experimental collaborations to study 
these observables in the next LHC analyses.

\acknowledgments{
G.C. acknowledges support from the MICINN, Spain, 
under contract FPA2013-44773-P. 
A.S.V. acknowledges support from Spanish Government 
(MICINN (FPA2015-65480-P)) and, 
together with F.C. and F.G.C., 
to the Spanish MINECO Centro de Excelencia 
Severo Ochoa Programme (SEV-2012-0249). 
F.G.C. thanks the Instituto de Fisica Teorica 
(IFT UAM-CSIC) in Madrid for warm hospitality.}

\end{document}